\documentstyle[12pt]{article}

\def\vbr{\vphantom{\sqrt{F_e^i}}}

\newcommand{\beq}{\begin{eqnarray}}
\newcommand{\eeq}{\end{eqnarray}}

\newcommand{\drawsquare}[2]{\hbox{%
\rule{#2pt}{#1pt}\hskip-#2pt
\rule{#1pt}{#2pt}\hskip-#1pt
\rule[#1pt]{#1pt}{#2pt}}\rule[#1pt]{#2pt}{#2pt}\hskip-#2pt
\rule{#2pt}{#1pt}}

\newcommand{\PSbox}[3]{\mbox{\rule{0in}{#3}\includegraphics{#1}\hspace{#2}}}
\newcommand{\Yfund}{\raisebox{-.5pt}{\drawsquare{6.5}{0.4}}}
\newcommand{\Ysymm}{\raisebox{-.5pt}{\drawsquare{6.5}{0.4}}\hskip-0.4pt%
        \raisebox{-.5pt}{\drawsquare{6.5}{0.4}}}
\newcommand{\Yasymm}{\raisebox{-3.5pt}{\drawsquare{6.5}{0.4}}\hskip-6.9pt%
        \raisebox{3pt}{\drawsquare{6.5}{0.4}}}

\newcommand{\jref}[4]{{\it #1} {\bf #2}, #3 (#4)}

\newcommand{\NPB}[3]{\jref{Nucl.\ Phys.}{B#1}{#2}{#3}}

\newcommand{\PLB}[3]{\jref{Phys.\ Lett.}{#1B}{#2}{#3}}

\newcommand{\PRD}[3]{\jref{Phys.\ Rev.}{D#1}{#2}{#3}}

\begin{document}
\begin{titlepage}
\begin{center}
{\hbox to\hsize{hep-th/9801207 \hfill   UCB-PTH-98/09}}
{\hbox to\hsize{               \hfill  LBNL-41341}}
{\hbox to\hsize{               \hfill  BU/HEP-98-04}}
{\hbox to\hsize{               \hfill UCSD-PTH-98/05}}
\bigskip

\bigskip

{\Large \bf  Gauge Theories with Tensors from \\ Branes and Orientifolds}
\bigskip

\bigskip

{\bf Csaba Cs\'aki$^{a,}$\footnote{Research Fellow, Miller Institute for
Basic Research in Science.},
Martin Schmaltz$^{b}$, Witold Skiba$^c$,
and \\
John Terning$^a$}\\

\smallskip

{\small \it $^a$ Department of Physics, University of California, Berkeley,
CA 94720}

and

{\small \it Theory Group, Lawrence Berkeley National Laboratory, Berkeley, CA
94720}

\smallskip

{\tt csaki@thwk5.lbl.gov, terning@alvin.lbl.gov}

\bigskip
{\small \it $^b$Department of Physics, Boston University,
Boston, MA 02215 }

\smallskip

{\tt schmaltz@abel.bu.edu}

\bigskip

{ \small \it $^c$Department of
Physics,

University of California at San Diego, La Jolla  CA 92093 }

\smallskip

{\tt skiba@einstein.ucsd.edu}

\bigskip

\vspace*{1cm}
{\bf Abstract}\\
\end{center}
We present brane constructions in Type IIA string theory for
${\cal N}=1$ supersymmetric $SO$ and $Sp$
gauge theories with tensor representations using an orientifold 6-plane.
One limit of these set-ups corresponds to ${\cal N}=2$ theories previously
constructed by Landsteiner and Lopez,
while a different limit yields ${\cal N}=1$
$SO$ or $Sp$ theories with a massless tensor and no superpotential.
For the $Sp$-type orientifold projection comparison with the field
theory moduli space leads us to postulate two new rules governing
the stability of configurations of D-branes intersecting
the orientifold.
Lifting one of our configurations to M-theory by finding the
corresponding curves, we re-derive the ${\cal N}=1$ dualities
for $SO$ and $Sp$ groups using semi-infinite D4 branes.

\bigskip

\bigskip

\end{titlepage}


\section{Introduction}
\setcounter{equation}{0}

A growing number of supersymmetric field theories have been realized as field
theories on the world volume of various brane configurations. While brane
dynamics provides information about field theory, one also learns about
properties of brane configurations from field theory knowledge. One considers
branes either in the context of perturbative string theory~[1-14] or embedded 
in M-theory or F-theory~[15-41].

Orientifold projection is a natural
way of obtaining $SO$ and $Sp$ theories on the world volume of branes,
and it can also yield field theories with two-index tensors. Many results
have been obtained by using an orientifold four plane~\cite{Nick,N2SoSp,
CW,DSB,AOT2,Gimon,JulieKen}. In this paper we
consider brane configurations with an orientifold projection
realized by an orientifold six plane (O6-plane).  These
configurations allow us to describe gauge theories which
have not been constructed previously using other brane configurations.

Our configurations can be understood as generalizations of the brane
set-up considered by Landsteiner and Lopez~\cite{LL} for ${\cal N}=2$
supersymmetric $SO$ and $Sp$ theories. Landsteiner and Lopez (LL)
considered D4 branes stretched between two parallel NS5 branes.
In between the NS5 branes there are D6 branes and an O6-plane.
The O6-plane amounts to a reflection involving the
space-time directions which are orthogonal to the orientifold.
As usual, the orientifold projection combines this spatial reflection with
a parity inversion on the string world sheet.

In this paper, we break the ${\cal N}=2$ supersymmetry of the LL
configuration by rotating the two NS5 branes. The rotation
corresponds to adding a mass term for the adjoint chiral
superfield in the field theory. A new interesting brane
configuration is obtained when the two NS5 branes are
rotated until they are parallel to the
orientifold, then the corresponding field theory has an
additional massless chiral superfield: a symmetric
(antisymmetric) tensor in the case of $SO$ ($Sp$).
We investigate the Coulomb and Higgs branches for all
our theories in detail and demonstrate agreement with
the field theory expectations.
For the $Sp$ configuration we find that two new rules
concerning the consistency of D branes and orientifolds are required.
The ``doubling rule" requires D branes which cross or
are parallel to the orientifold to come in pairs (thus
explaining why $Sp$ theories only exist for even numbers of
colors and flavors), and the ``s'-rule" which is
similar to the ``s-rule" of Hanany and Witten~\cite{HananyWitten},
and which is necessary
to reproduce the dimension of the Higgs branch of the field
theory moduli space correctly.

Finally, we also consider brane configurations corresponding to
$SO$ and $Sp$ theories with only fundamental matter fields. We
explicitly construct the corresponding M-theory curves
and use them to re-derive Seiberg's duality for $SO$ and $Sp$ gauge theories.
This derivation is similar to the derivation of Seiberg's
duality using an O4 plane in Ref.~\cite{CW}.

The paper is organized as follows. In the next section, we describe the basic
brane configuration and determine the corresponding low-energy field theory.
In Section 3, we study the Higgs branch of theories with
flavors. In Section 4, we derive duality for $SO$ and $Sp$ with fundamentals
in the context of string theory by considering brane motions. We discuss the
embedding of this set-up in M-theory and present curves describing our
brane configuration in Section 5. Finally, we conclude in the last section.

Certain elements of our paper have appeared very recently in
Refs.~\cite{kutasov,Karch}. In particular, Brunner, Hanany, Karch and
Lust~\cite{Karch}
mention the brane configurations of Sec.~\ref{sec:setup}. The counting of the
${\cal N}=2$ moduli spaces of Section~\ref{sec:higgs} together with a slightly
different statement of the s'-rule of Section~\ref{sec:higgs}
(with the same physical
consequences) appeared in the work of Elitzur, Giveon, Kutasov and
Tsabar~\cite{kutasov}.


\section{The Brane Set-up\label{sec:setup}}
\setcounter{equation}{0}

\begin{figure}
\PSbox{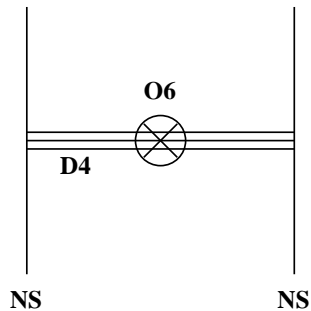 hscale=70 vscale=70 hoffset=150  voffset=0}{13.7cm}{4.5cm}
\caption{The brane configuration in Type IIA string theory
giving rise to the pure ${\cal N}=2$ $SO$ or $Sp$ theories.
The $\otimes$ denotes the O6-plane which is perpendicular to the
NS5 branes. \label{LL}}
\end{figure}

In this section we consider brane configurations
in Type IIA string theory which give rise to $SO$ and $Sp$
gauge groups. Our starting point is the ${\cal N}=2$ configuration of
Ref.~\cite{LL} presented in Fig.~\ref{LL}. In this configuration we have
an O6-plane in the $(x_0,x_1,x_2,x_3,x_7,x_8,x_9)$ directions, which thus acts
as a mirror\footnote{We present
all of our brane configurations as embedded into the double
cover of the orientifolded space.} in $(x_4,x_5,x_6)$.
In addition, we have two NS5 branes 
in the $(x_0,x_1,x_2,x_3,x_4,x_5)$ directions
which are mirror images of each other under the orientifold projection.
There are also $N$ D4 branes in the $(x_0,x_1,x_2,x_3,x_6)$
directions connecting
the two NS5 branes. This theory, as discussed in Ref.~\cite{LL},
corresponds to ${\cal N}=2$ $SO(N)$ theory or
$Sp(N)$ theory depending on the orientifold charge.
First we will discuss the ${\cal N}=1$ $SO(N)$ theory obtained
from this ${\cal N}=2$ set-up by rotation of the NS5 branes. Then we repeat the
discussion for $Sp$ groups. The analysis for the $Sp$ theories is
very similar to the case of $SO$ groups, with the exception of an
important subtlety which we discuss in detail. Throughout this
paper, we use the complex coordinates $v=x_4+ix_5$ and
$w=x_8+ix_9$. With this notation, in the ${\cal N}=2$ theory the
NS5 branes are at a point in the $w$ plane and fill out the $v$ plane,
while the O6 is at the origin of $v$ and fills out $w$.

\subsection{The $SO(N)$ Theories}

\begin{figure}
\PSbox{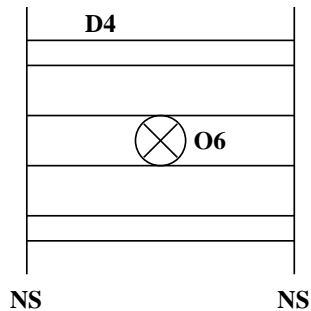 hscale=70 vscale=70 hoffset=150  voffset=0}{13.7cm}{4.5cm}
\caption{The moduli space of the ${\cal N}=2$ $SO(N)$ theory. Note that the
D4 branes have to move in pairs away from the O6-plane.
\label{SOmoduli}}
\end{figure}

The ${\cal N}=2$ $SO(N)$ theory is given in Fig.~\ref{LL}. The moduli space 
of this
theory corresponds to giving expectation values to the adjoint of the
$SO(N)$ (which is an antisymmetric tensor). In the brane language
this corresponds to sliding the D4 branes between the two parallel
NS5 branes. At a generic point of the moduli space the adjoint VEV
breaks $SO(N)$ to $U(1)^r$, where $r=N/2$ for $N$ even and $r=(N-1)/2$
for $N$ odd, which just corresponds to sliding all D4 branes apart
from each other as illustrated in Fig.~\ref{SOmoduli}. Due to the
O6 projection the D4 branes have to slide between the NS5 branes
in pairs in the opposite directions, thus for $N$ even we get
as a dimension of the moduli space $N/2$. For $N$ odd one of the
D4 branes is stuck at the O6, and the number of moduli is
given by $(N-1)/2$.

The ${\cal N}=2$ $SO(N)$ theory has an anomalous $U(1)_R$ symmetry, under which
the adjoint field (the antisymmetric tensor of $SO(N)$) carries charge
two. The $U(1)_R$ can be identified with rotations of the $v$ (45) plane,
$v \rightarrow e^{i\theta} v$, we call this symmetry $R_v$. Thus, the $R_v$ 
charge of the adjoint is two.

\begin{figure}
\PSbox{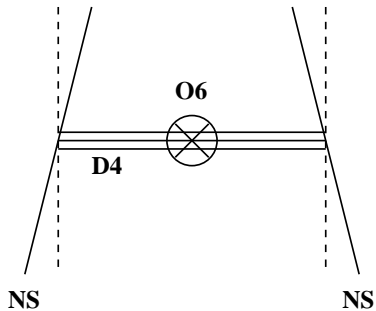 hscale=70 vscale=70 hoffset=150  voffset=0}{13.7cm}{4.5cm}
\caption{The brane configuration obtained by the rotation of the
NS5 branes of the ${\cal N}=2$ $SO(N)$ theory.
\label{rotate}}
\end{figure}

Let us now rotate the NS5 branes slightly out of the $v$ plane into the
$w$ plane (Fig.~\ref{rotate}).
Since the configuration must remain symmetric
under the orientifold projection, both NS5 branes have to be
rotated, but in opposite directions. As a result of the rotation, the
D4 branes are fixed at the origin, they can not slide between
the NS5 branes anymore. Thus the moduli space of the theory is lifted.
This is exactly analogous to what happens in the ${\cal N}=2$ $SU(N)$ theories
when one of the NS5 branes is rotated~\cite{HOO,Witten2}. 

In the field theory a mass for the adjoint chiral superfield is
generated, breaking ${\cal N}=2$ to ${\cal N}=1$,
and lifting the Coulomb branch of the theory. The
antisymmetric tensor $A$ gets a small mass $W=\mu A^2$ when the
two NS5 branes are slightly rotated. The $\mu \to 0$ limit corresponds to the
${\cal N}=2$ theory discussed before, and for generic $\mu$ we obtain
pure ${\cal N}=1$ $SO(N)$ gauge theory with no massless chiral superfields.
This theory has no moduli space.

\begin{figure}
\PSbox{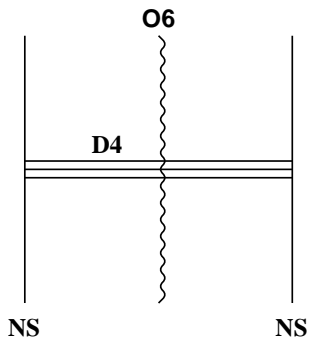 hscale=70 vscale=70 hoffset=150
voffset=0}{13.7cm}{4.5cm}
\caption{The brane configuration obtained in the $\mu \to \infty$ limit,
which gives
rise to ${\cal N}=1$ $SO(N)$ theory with a massless symmetric tensor, or
$Sp(N)$ with an antisymmetric tensor.
\label{puretensor}}
\end{figure}

However, we can observe that there is another limit of the rotated
brane configuration where the $SO(N)$ theory does have a moduli space.
If the angle of rotation is $\pi /2$ the two NS5 branes become parallel
to each other and the O6-plane, and the D4 branes can
slide between the NS5 branes again (Fig.~\ref{puretensor}).
The emerging moduli space of the brane configuration must correspond
to a field whose mass goes to zero in the field theory. We conjecture
that this field transforms as a symmetric tensor under the $SO(N)$
gauge group.\footnote{In the classical
picture the trace of the symmetric tensor of $SO(N)$ appears as a
modulus as well.
This degree of freedom is presumably frozen by
an infrared divergence of the quantum theory, similar to the trace
of the adjoint of the ${\cal N}=2$ $SU(N)$ theories~\cite{Witten}.}. 
We now present four different arguments in support of this
conjecture.

\begin{figure}
\PSbox{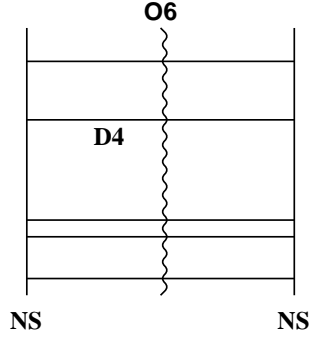 hscale=70 vscale=70 hoffset=150
voffset=0}{13.7cm}{4.5cm}
\caption{The moduli space of the ${\cal N}=1$ $SO(N)$ theory with a symmetric
tensor. If all D4 branes move apart the gauge group is completely broken.
\label{N=1SOmoduli}}
\end{figure}

\begin{figure}
\PSbox{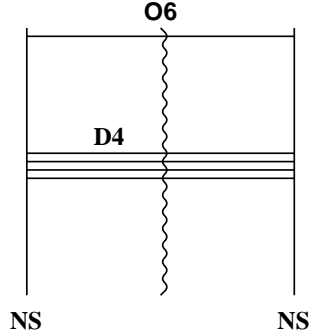 hscale=70 vscale=70 hoffset=150
voffset=0}{13.7cm}{4.5cm}
\caption{The brane configuration which corresponds to the
breaking $SO(N)\to SO(N-1)$.
\label{SObreaking}}
\end{figure}

First we compare the moduli space of the field theory of $SO(N)$ with a
symmetric tensor to that of the brane picture. In the brane picture
the moduli space corresponds to the sliding of the D4 branes along the
parallel NS5 branes (see Fig.~\ref{N=1SOmoduli}).
This obviously has $N$ moduli,
and since a single D4 brane in the presence of an O6 corresponds to
``$SO(1)$'', this means that on a generic point of the moduli space
the $SO(N)$ is completely broken. Let us now consider the unbroken
group if we move away only one D4 brane (Fig.~\ref{SObreaking}).
Then we still have
$N-1$ D4 branes on top of each other and thus this breaking
corresponds to $SO(N)\to SO(N-1)$. This brane picture is in complete
agreement with the field theory result. In the field theory,
giving a VEV
\begin{displaymath} 
V
\left( \begin{array}{ccccc}
1 & \\
& 1 & \\
& & \ddots \\
& & & 1 \\
& & & & -(N-1) \end{array} \right)
\end{displaymath}
to the symmetric tensor breaks the gauge group to $SO(N-1)$, and there is
a massless symmetric tensor of the $SO(N-1)$ group remaining. This
can be identified with the above brane motion where one of the D4 branes is
sliding away from the remaining $N-1$. In general, the symmetric tensor
$S$ of $SO(N)$ can completely break the $SO(N)$ group, which corresponds
to moving all D4 branes apart from each other. Finally, the independent gauge
invariant operators are given by ${\rm Tr} S,{\rm Tr} S^2,
\ldots , {\rm Tr} S^N$\cite{DM}, and thus the number of moduli
agrees with the brane prediction. In summary, we find complete
agreement between the field theory and the brane moduli space.
 Note, that the antisymmetric tensor
(adjoint) would not give the right dimension of moduli space, since an
${\cal N}=2$ $SO(2M)$ theory has $M$ moduli.

\begin{figure}
\PSbox{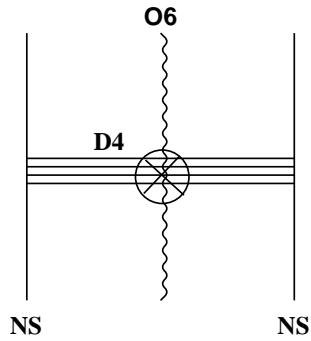 hscale=70 vscale=70 hoffset=150
voffset=0}{13.7cm}{4.5cm}
\caption{The brane configuration corresponding to an ${\cal N}=1$ $U(N)$
with a symmetric and a conjugate symmetric tensor. The $\otimes$ corresponds
to a third NS5 brane perpendicular to the other two NS5 branes and the
O6.
\label{SU}}
\end{figure}

\begin{figure}
\PSbox{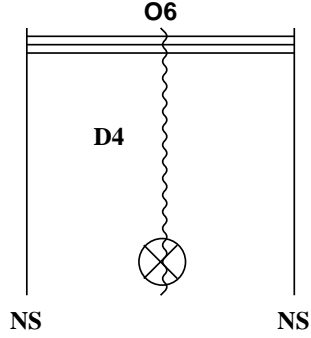 hscale=70 vscale=70 hoffset=150
voffset=0}{13.7cm}{4.5cm}
\caption{The brane configuration corresponding to breaking
$U(N)\to SO(N)$ by giving an expectation value to the symmetric tensor.
\label{SUhiggs}}
\end{figure}

As a second piece of evidence for establishing that the brane configuration
of Fig.~\ref{puretensor} corresponds to $SO(N)$ with a symmetric tensor
consider the brane set-up of Fig.~\ref{SU} (we will discuss this brane
configuration in more detail in Section~\ref{SUsection}).
Here we have three NS5 branes,
two of which are parallel to the O6-plane and are thus in the
$(x_0,x_1,x_2,x_3,x_8,x_9)$
 directions, while the third NS5 brane is at the same point as the
O6 in the $x_6$ direction, but is perpendicular to it, since it is
in the $(x_0,x_1,x_2,x_3,x_4,x_5)$ 
directions. This theory corresponds to ${\cal N}=1$ $U(N)$  
theory with a symmetric tensor and its conjugate.  We can however
move the D4 branes away from the middle NS5 brane by sliding them
between the parallel NS5 branes, and the resulting brane
configuration is that of Fig.~\ref{puretensor}
(see Fig.~\ref{SUhiggs}). In the field theory this corresponds to the
higgsing of $U(N)$ to $SO(N)$ by a VEV for a symmetric tensor which
leaves one massless symmetric tensor transforming under the $SO(N)$
as expected from the brane picture.

As a third piece of evidence consider the argument in Section 2 of
the work of Landsteiner and Lopez~\cite{LL}. LL argue that $N$ D4 branes
in an orientifold background give rise to two kinds of matter
multiplets: an adjoint and a two-index tensor other than the
adjoint. The boundary conditions imposed on the states on the
D4 branes by the transverse NS5 branes project out some of
these states. Which states are projected out depends on the orientation
of the NS5 branes. If they 
are perpendicular to the O6 the tensor is projected out, thus
leaving an ${\cal N}=2$ theory. If the NS5 branes are parallel to the O6 the  
adjoint
and part of the tensor is projected out leaving $SO(N)$ with a
symmetric tensor.

\begin{figure}
\PSbox{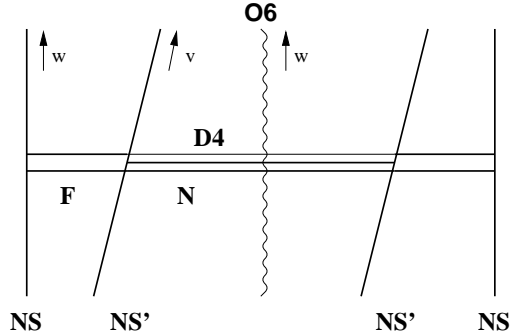 hscale=70 vscale=70 hoffset=130
voffset=0}{13.7cm}{4.5cm}
\caption{The brane configuration which corresponds to an ${\cal N}=2$
$SO(N)$ theory with $F$ flavors and with
the $U(F)$ flavor symmetry gauged.
\label{product}}
\end{figure}

As final piece of evidence, consider the brane configuration of
Fig.~\ref{product}. There, the theory between the middle two NS5 branes
is an ${\cal N}=2$ $SO(N)$ theory, and due to the additional $F$ D4 branes
we get $F$ hypermultiplets of this ${\cal N}=2$ theory. Because the additional
D4 branes end on two extra NS5 branes at the left and the right of the
set-up, the $U(F)$ flavor symmetry is gauged.
Thus, the field theory corresponding to Fig.~\ref{product} is given by
\[
\begin{array}{ccc}
& SO(N) & U(F) \\
\Phi & \Yasymm=Adj & 1 \\
Q& \Yfund & \Yfund \\
\bar{Q} & \Yfund & \overline{\Yfund} \end{array}.
\]
The superpotential of the theory is
\beq
 W= Q\Phi \bar{Q}.
\eeq

\begin{figure}
\PSbox{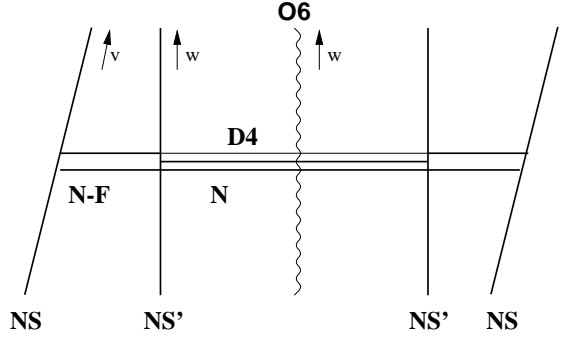 hscale=70 vscale=70 hoffset=130
voffset=0}{13.7cm}{4.5cm}
\caption{The brane configuration obtained by dualizing the $SU(F)$ group
in Fig.~\protect\ref{product}.
\label{dualproduct}}
\end{figure}

Now let us apply Seiberg's ${\cal N}=1$ duality to the $U(F)$ group
(assuming that $F<N$). In the field theory this results in the theory

\[
\begin{array}{ccc}
& SO(N) & U(N-F) \\
\Phi & \Yasymm=Adj & 1 \\
q& \Yfund & \Yfund \\
\bar{q} & \Yfund & \overline{\Yfund} \\
M_S & \Ysymm & 1 \\
M_A & \Yasymm & 1
\end{array},
\]
where $M_A$ is the antisymmetric part of the meson field, while
$M_S$ is the symmetric part under the gauged $SO(N)$ subgroup
of the original $U(N)$ flavor symmetry.
The superpotential after the duality transformation is
\beq
 W=M_A \Phi +(q\bar{q})_S M_S + (q\bar{q})_A M_A,
\eeq
which gives mass to the original adjoint $\Phi$ of $SO(N)$ together with
the antisymmetric part of the meson $M_A$. Thus we are left with an
${\cal N}=1$ $SO(N)$ theory with a symmetric tensor and $F-N$  
flavors\footnote{Note 
that the additional $SO(N)$ singlet which is the trace of the 
symmetric tensor remains in the classical theory  in
this construction as well.}.
The corresponding brane configuration is given in Fig.~\ref{dualproduct},
from which one can see that the configuration of Fig.~\ref{puretensor} must
correspond to $SO(N)$ with a massless symmetric tensor.

\begin{figure}
\PSbox{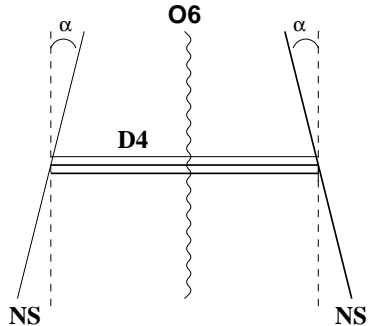 hscale=70 vscale=70 hoffset=150  voffset=0}{13.7cm}{4.5cm}
\caption{The general rotated brane configuration in Type IIA string theory
giving rise to ${\cal N}=1$ $SO$ or $Sp$ theories.\label{setup}}
\end{figure}

Thus, we have established that the brane configuration of Fig.~\ref{puretensor}
corresponds to an ${\cal N}=1$ $SO(N)$ theory with a symmetric tensor. This  
theory
has an anomalous $R$-symmetry under which the tensor has charge two.
This symmetry can be identified with the rotations of the
$w$ plane (89), $R_w$, and the $R_w$ charge of the tensor is two.
In the limit of Fig.~\ref{puretensor} which gives rise to the
$SO(N)$ theory with a symmetric tensor, the mass of the adjoint is
infinitely large, $\mu \to \infty$. Performing a small rotation
around the $\frac{1}{\mu}=0$ limit we expect that the mass of the
symmetric tensor is proportional to $\frac{1}{\mu}$. This suggests that
the configuration with the general rotation angle in Fig.~\ref{setup}
corresponds to
${\cal N}=1$ $SO(N)$ theory with an adjoint $A$ and a symmetric tensor $S$, and  
a
superpotential
\beq
W=\mu A^2 +\frac{1}{\mu} S^2.
\eeq
The anomalous R-symmetries of the fields are given by
\[
\begin{array}{ccc}
& R_v & R_w \\
A & 2 & 0 \\
S & 0 & 2 \\
\mu & -2 & 2 \end{array}, \]
showing that the $R_v$ and $R_w$ charges of $\mu$ are consistent with
our interpretation of $\mu$ as the adjoint mass and the inverse mass
of the symmetric tensor.

\subsection{The $Sp(2N)$ Theories}

We now consider the same brane set-up as in the previous section except
with negative O6 charge.
The set-up of Fig.~\ref{LL} gives rise to an ${\cal N}=2$ $Sp(2N)$
theory~\cite{LL}.
An important new element compared to the $SO(N)$ case is recognized by
considering the fact that there is no $Sp(N)$ group for $N$ odd.
This leads us to postulate the

{\it doubling rule: a single D4 brane passing
through the O6-plane is not invariant under an $Sp$-type O6 projection,
therefore the D4 branes are necessarily paired.}

\begin{figure}
\PSbox{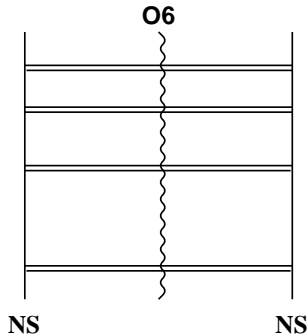 hscale=70 vscale=70 hoffset=150
voffset=0}{13.7cm}{4.5cm}
\caption{The moduli space of an $Sp(2N)$ theory with an antisymmetric
tensor. Due to the orientifold projection the D4 branes can slide only in
pairs between the NS5 branes.
\label{Spmoduli}}
\end{figure}

The moduli space of the ${\cal N}=2$ $Sp(2N)$ theory is $N$ dimensional,
corresponding to the breaking $Sp(2N)\to U(1)^N$ by the adjoint (which is the
symmetric tensor) of $Sp(2N)$ (see Fig.~\ref{SOmoduli}).
Again, the rotation of the
NS5 branes of Fig.~\ref{LL} gives  mass $\mu S^2$ to the adjoint.
The $\mu \to \infty$ limit
($\pi/2$ rotation) again gives a theory with a moduli space, corresponding to
${\cal N}=1$ $Sp(2N)$ theory with an antisymmetric tensor $A$. An important
difference compared to $SO(N)$ appears when identifying the
moduli space of the $Sp(2N)$ theory with an antisymmetric
tensor. Due to the doubling rule the D4 branes  must
move in pairs. 

A generic point on the moduli space is depicted in Fig.~\ref{Spmoduli}.
Note that since two D4 branes are always on top of each other, the
general unbroken subgroup is predicted to be $SU(2)^N$. 
In field theory the most general expectation value 
of the antisymmetric tensor is 
\[ A=\left( \begin{array}{cccc}
\omega_1 \\
& \omega_2 \\
& & \ddots \\
& & & \omega_N \end{array} \right) \otimes i\tau_2 \]
which higgses $Sp(2N)$ to $SU(2)^N$, in agreement with the brane picture.
Another way of seeing the agreement is to note that there are exactly $N$
independent gauge invariant operators of $Sp(2N)$ with an
antisymmetric tensor $A$, given by
${\rm Tr} AJ,{\rm Tr} (AJ)^2,\ldots ,{\rm Tr} (AJ)^N$ where
$J$ is the symplectic invariant~\cite{Sp}.

In all other respects the treatment of the $Sp(2N)$ theories is
in complete analogy with the
$SO(N)$ theories. The model obtained for a general rotation angle is
$Sp(2N)$ with an adjoint (the symmetric tensor $S$) and an antisymmetric tensor
$A$, with mass terms
\beq
W=\mu S^2 +\frac{1}{\mu} A^2,\eeq
where $\mu \to 0$ for the ${\cal N}=2$ theory, and $\mu \to \infty$ for the  
case when
the NS5 branes and the O6-plane are parallel. The charges under the
anomalous symmetries are:

\[
\begin{array}{ccc}
& R_v & R_w \\
S & 2 & 0 \\
A & 0 & 2 \\
\mu & -2 & 2 \end{array}. \]

\subsection{$U(N)$ Theories with Symmetric or Antisymmetric
Tensors \label{SUsection}}

\begin{figure}
\PSbox{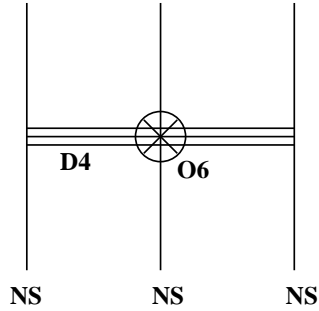 hscale=70 vscale=70 hoffset=150
voffset=0}{13.7cm}{4.5cm}
\caption{The brane configuration of Ref.~\protect\cite{LL} giving rise
to ${\cal N}=2$ $U(N)$ theories with a symmetric or an antisymmetric flavor.
The $\otimes$ denotes the O6-plane.\label{SUsymmN=2}}
\end{figure}

\begin{figure}
\PSbox{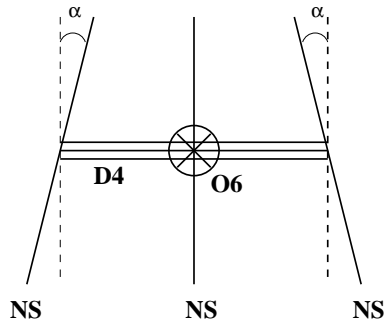 hscale=70 vscale=70 hoffset=150
voffset=0}{13.7cm}{4.5cm}
\caption{The brane configuration giving rise
to ${\cal N}=1$ $U(N)$ theories with a symmetric or an antisymmetric flavor
and a tree-level superpotential for them.
\label{LLrot}}
\end{figure}

In this section we consider brane configurations corresponding to
${\cal N}=1$ $U(N)$ theories\footnote{Quantum mechanically, the Abelian
$U(1)$ factor of the $U(N)$ is not dynamical~\cite{Witten}.}
with a symmetric flavor ($\Ysymm  
+\overline{\Ysymm}$)
or an antisymmetric flavor ($\Yasymm +\overline{\Yasymm}$). The starting point
is the ${\cal N}=2$ $U(N)$ configuration of Landsteiner and Lopez~\cite{LL}  
given in
Fig.~\ref{SUsymmN=2}. The O6-plane is in the $(x_0,x_1,x_2,x_3,x_7,x_8,x_9)$
plane ($w$), while
the NS5 branes are in the $(x_0,x_1,x_2,x_3,x_4,x_5)$ ($v$) plane.
Depending on the charge of the O6-plane the theory is ${\cal N}=2$ $U(N)$  
theory
with a symmetric or an antisymmetric flavor.
Rotation of the two outside NS5 branes gives a mass to the
$U(N)$ adjoint breaking ${\cal N}=2$ to ${\cal N}=1$ (Fig.~\ref{LLrot}). The
superpotential of the theory is
\beq
W= \mu A^2 +TA\bar{T},
\eeq
where $A$ is the $U(N)$ adjoint, and $T, \bar{T}$ are the symmetric
or antisymmetric tensors of $U(N)$ and their conjugates. The second piece
of the superpotential is the usual ${\cal N}=2$ superpotential term.
Integrating out the massive adjoint field one can see that the configuration
of Fig.~\ref{LLrot} corresponds to an ${\cal N}=1$ $U(N)$ theory
with a symmetric or antisymmetric flavor and a tree-level superpotential term
\beq
W\propto \frac{1}{\mu} (T\bar{T})^2.
\eeq
The configuration obtained in the limit when $\mu \to \infty$ corresponds to
the ${\cal N}=1$ theory with $T,\bar{T}$ and no tree-level superpotential. This
happens, when the two outside NS5 branes are parallel to each other,
and to the O6-plane. The fact that the superpotential is vanishing in this
limit can also be seen from the brane picture, where now the D4 branes can
slide between the two outside NS5 branes, corresponding to expectation values
of $T,\bar{T}$. This limiting case is depicted in Fig.~\ref{SU}, while
the moduli space in this limit is illustrated in Fig.~\ref{SUhiggs}.
This is the brane motion we have used to establish that
the configuration of Fig.~\ref{puretensor} corresponds to ${\cal N}=1$
$SO$/$Sp$ theories with a tensor and no superpotential.
Note that the moduli space of this $U(N)$ theory is exactly the
same as the moduli space of the ${\cal N}=1$ $SO$/$Sp$ theories with a
massless tensor of Fig.~\ref{puretensor}, which agrees with the
brane picture, since the presence of the middle NS5 brane is
irrelevant for that.

\begin{figure}
\PSbox{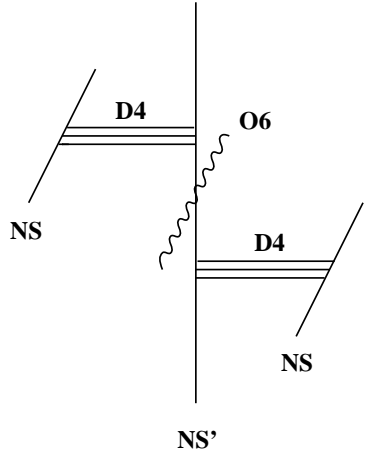 hscale=70 vscale=70 hoffset=150
voffset=0}{13.7cm}{6.0cm}
\caption{The brane configuration giving rise
to ${\cal N}=1$ $U(N)$ theories with a massive symmetric or an antisymmetric  
flavor.
The mass of the tensors corresponds to the separation of the D4 branes
along the middle NS5 brane.
\label{massivetensor}}
\end{figure}

Finally, we remark that the brane configuration corresponding to the
${\cal N}=1$ $U(N)$ theory with a massive symmetric or antisymmetric flavor
is the one given in Fig.~\ref{massivetensor}. The mass of the tensors
corresponds to the separation of the D4 branes along the middle
NS5 brane.

\section{The Higgs Branch\label{sec:higgs}}
\setcounter{equation}{0}

In this section we are introducing $F$ D6 branes on each side of the
O6-plane parallel to it. This will enable us
to describe the $SO/Sp$ theories with tensors and $F$
additional flavors. These theories, depending on the actual field content
and superpotential usually have Coulomb branches, Higgs branches and
mixed Coulomb-Higgs branches. We fill focus only on the pure Higgs
branches of these theories, thus setting the tensor expectation value
to zero.
We are going to compare the dimension of the Higgs branch
of the field theory
when the gauge group is completely higgsed to the brane prediction.
The moduli space appears in the brane picture because the D4 branes can break
up between the D6 branes and the separate pieces can slide individually
between the D6 branes.

In order to count the dimension of the moduli space correctly, the
following counting rules (as stated in Ref.~\cite{Kutasov}) have to be
taken into account:

-  a D4 brane  between two D6 branes has two complex degrees of freedom
(there are three real scalars corresponding to the  three directions
the D4 brane can move in and there is an additional scalar from the
Kaluza-Klein compactification of the gauge field in the direction of the
D4 brane).

- a D4 brane between a D6 brane and a parallel NS5 brane has one complex
degree of freedom (there are two real scalar fields corresponding to the
two directions the D4 brane can move in).

In addition to these counting rules, for a D4 brane stretched between
an NS5 brane and a D6 brane perpendicular to the NS5 the following rule
has to be obeyed~\cite{HananyWitten}:

{\it s-rule: there can be only one D4 brane suspended between a D6 brane and a
perpendicular NS5 brane.}

There are several pieces of evidence in favor of the s-rule.
First of all, the neat fit with the field theory which we will
show in the following subsections is only possible with the s-rule
in place. Then, if there was more than one D4 brane stretched between
the NS5 and the perpendicular D6 they would necessarily be on top
of each other, a singular situation which presumably breaks
supersymmetry~\cite{Kutasov}. Further evidence for the s-rule
can be derived from M-theory. One can write down the M-theory curve
for the given brane configuration and count the number of complex spheres 
which decouple from the curve and are free to slide along the D6 branes.
This argument has been given in Ref.~\cite{HOO}, the result
is in exact agreement with the s-rule.

\subsection{The $SO(N)$ Theories}

Using the counting rules presented above we can check the dimension of the
moduli space of the $SO$ theories of the previous section. First we
consider the ${\cal N}=2$ $SO(N)$ theory.

\begin{figure}
\PSbox{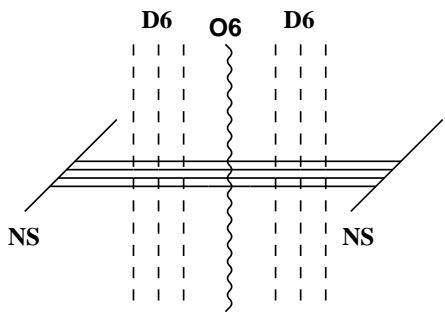 hscale=70 vscale=70 hoffset=150
voffset=0}{13.7cm}{4.5cm}
\caption{The ${\cal N}=2$ $SO(N)$ theory
and $F$ hypermultiplets.
\label{N=2SOD6}}
\end{figure}

\begin{figure}
\PSbox{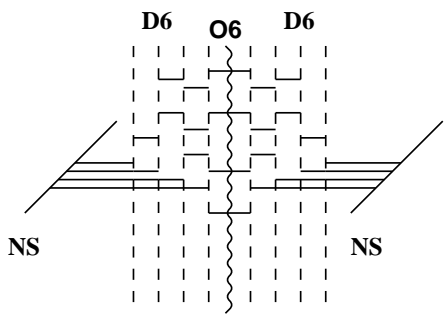 hscale=70 vscale=70 hoffset=150
voffset=0}{13.7cm}{4.5cm}
\caption{The moduli space of the Higgs branch of the ${\cal N}=2$ $SO(N)$  
theory
and $F$ hypermultiplets.
\label{N=2SO}}
\end{figure}

The set-up of this theory is given in Fig.~\ref{N=2SOD6}. The D6 branes fill
out the $(x_0,x_1,x_2,x_3,x_7,$ $x_8,x_9)$ directions
just as the O6-plane, and are at definite points in $(x_4,x_5,x_6)$. 
The NS5 branes
fill out the $(x_0,x_1,x_2,x_3,x_4,x_5)$  directions. If we want to
identify the Higgs branch, we have to consider the case when the
D6 branes are at the origin of $(x_4,x_5)$ 
(massless flavors). In this case, the
D4 branes can break up and slide between the D6 branes. This will be the
brane picture corresponding to the Higgs branch of the field theory.
We want to find the maximal size of this Higgs branch.
Taking into account the s-rule between the D6 and the NS5 branes one obtains
for the  Higgs branch a brane configuration as depicted in Fig.~\ref{N=2SO}.

Thus, the brane prediction for the size of the moduli space is
\beq
\sum_{i=1}^N 2i +(F-N)2N= 2FN-N^2+N.
\eeq
The field theory answer is exactly the same, which can be obtained by
considering that we have $2FN$ degrees of freedom from the quarks,
but there are $N(N-1)/2$ D-flatness conditions.
However, due to the superpotential
$QA\bar{Q}$ of the ${\cal N}=2$ theory, where $Q$,$\bar{Q}$ are the $SO(N)$  
vectors
forming one hypermultiplet and $A$ is the $SO(N)$ adjoint, there are
$N(N-1)/2$ F-flatness  conditions that these VEVs have to
satisfy. Thus, the field theory result exactly matches the above brane
prediction.

\begin{figure}
\PSbox{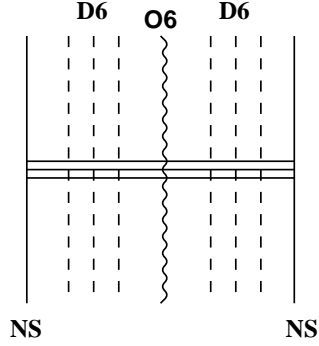 hscale=70 vscale=70 hoffset=150
voffset=0}{13.7cm}{4.5cm}
\caption{The basic set-up after including D6 branes parallel to the
O6-plane. This will give rise to $SO$/$Sp$ with a tensor and $F$ flavors.
\label{sixbranes}}
\end{figure}

\begin{figure}
\PSbox{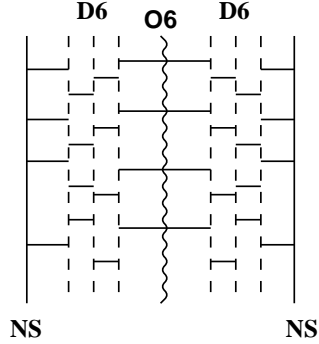 hscale=70 vscale=70 hoffset=150
voffset=0}{13.7cm}{4.5cm}
\caption{The moduli space of the Higgs branch of the $SO(N)$ theory
with a symmetric tensor and $F$ massless flavors.
\label{SOHiggs}}
\end{figure}

Let us now consider the other limiting case with a massless tensor,
the ${\cal N}=1$ $SO(N)$ theory with a symmetric tensor and $F$ flavors. The
brane set-up is given in Fig.~\ref{sixbranes}. Now the NS5 and the
D6 are parallel so the s-rule does not apply. Therefore the brane
configuration for the Higgs branch is as in Fig.~\ref{SOHiggs}.

Using these rules, the result for $SO(N)$ theories with $F$ flavors ($2F$
fields in the vector representation) and a symmetric tensor is given by
\beq
 N+\underbrace{2N+2N+\ldots +2N}_{F}=N+2FN.
\eeq
In the field theory, there are $2FN$ complex degrees of freedom from the
$2F$ fields in the vector representation, $N(N+1)/2$ from the
symmetric tensor, but there are $N(N-1)/2$ D-term conditions.
Since there is no tree-level superpotential term present,
the moduli space is
\beq
 2FN+N(N+1)/2-N(N-1)/2=2FN+N
\eeq
dimensional, again exactly in agreement with the brane prediction.

\subsection{The $Sp(2N)$ Theories}

The counting of the moduli space dimension for the $Sp(2N)$ theories is very
similar to the case of $SO$ theories presented in the previous section, with
the exception of two interesting subtleties arising from the negative
projection of the orientifold. First, we have to take into account the
doubling rule presented above (a single D4 brane passing through the
O6-plane with negative charge is not invariant under the O6 projection)
forcing the D4 branes which pass through the O6 to be paired.
Furthermore, we find that in order to
obtain the correct dimension of the moduli space the following
generalized version of the s-rule has to hold:

{\it s'-rule: when two or more D4 branes are forced to pass through
a perpendicular D6 brane at the same point, only one of the
D4 branes is allowed to break and end on the D6 brane.}

In the original s-rule considered by Hanany and Witten the D4 branes
were forced to be incident at a single point by their boundary condition
on a perpendicular NS5 brane. In the case of a parallel NS5 brane
the D4 branes are not forced to be coincident because they can be
separated by sliding them between the parallel D6 and NS5 branes. Thus
in that case the s'-rule does not apply. In the case at hand, the
D4 branes are forced to come in pairs by the orientifold projection,
and therefore only one of the two D4 branes can break on the D6 brane.
Note that our s'-rule implies the counting rule given in
Ref.~\cite{kutasov}.
The arguments in favor of the more general s'-rule rule are similar
to those given for the usual s-rule. If more than one D4 brane
were to end on the same D6 brane, then they would be necessarily
on top of each other, which is a singular situation and is likely
to break supersymmetry. We speculate that the s'-rule can also be
derived by going to M-theory and counting the number of spheres
which can decouple from the rest of the curve and slide along the
sixbranes as done for the case of the
s-rule in Ref.~\cite{HOO}. Finally, the neat fit between the field
theory and string theory results presented below has to be considered
a strong argument in favor of the s'-rule.

\begin{figure}
\PSbox{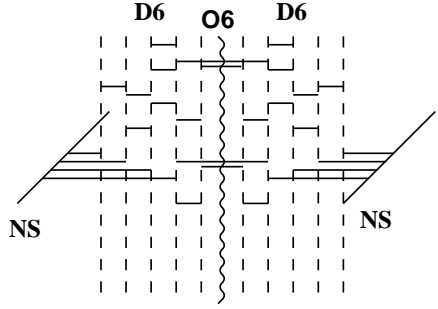 hscale=70 vscale=70 hoffset=150
voffset=0}{13.7cm}{4.5cm}
\caption{The moduli space of the Higgs branch of the ${\cal N}=2$ $Sp(2N)$  
theory
and $F$ hypermultiplets.
\label{SpN=2}}
\end{figure}

With this s'-rule in hand we can now
calculate the size of the moduli
space in the $Sp$ theories. First consider the
${\cal N}=2$ $Sp(2N)$ theories.
The brane configuration respecting the s'-rule for the
completely broken gauge group
is given in Fig.~\ref{SpN=2}.
Thus the dimension of the moduli space for $Sp(2N)$ is predicted to be
\beq
\sum_{i=1}^{2N} 2 i +(F-2N-2)4N +4N=4FN-4N^2-2N.
\eeq
In the field theory one has $4FN$ degrees of freedom from the quarks, but there
are $2N(2N+1)/2$ D-flatness conditions. Due to the superpotential term of
${\cal N}=2$ theories there are $2N(2N+1)/2$ additional
F-flatness conditions, exactly reproducing
the brane prediction $4FN-4N^2-2N$.

\begin{figure}
\PSbox{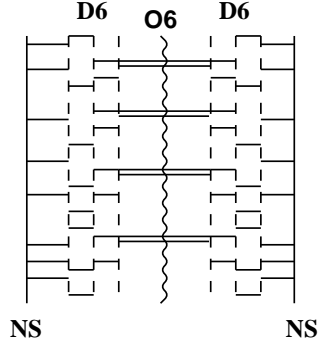 hscale=70 vscale=70 hoffset=150
voffset=0}{13.7cm}{4.5cm}
\caption{The moduli space of the Higgs branch of the $Sp(N)$ theory
with an antisymmetric tensor and $F$ massless flavors.
\label{SpHiggs}}
\end{figure}

Next consider the ${\cal N}=1$ $Sp(2N)$ theory with an
antisymmetric tensor. The most general brane configuration
respecting the s'-rule is
given in Fig.~\ref{SpHiggs}. The size of the moduli space for the $Sp(2N)$
theory is thus predicted to be
\begin{equation}
 2N+\underbrace{4N+4N+\ldots +4N}_{F-2}+2N+2N=6N+4N(F-2)=4NF-2N.
\end{equation}
The field theory result is in complete agreement. The
$2F$ quarks give rise to $2F\, 2N$ complex degrees of freedom, the
antisymmetric tensor to $2N(2N-1)/2$, but there are $2N(2N+1)/2$
D-flatness conditions, resulting in $4NF+N(2N-1)-N(2N-1)=4NF-2N$
degrees of freedom. Thus, we conclude that the counting
of moduli space is correctly reproduced in every example once the
s'-rule is taken into account.

\newpage

\section{${\cal N}=1$ Duality from String Theory}
\setcounter{equation}{0}
\subsection{Duality of ${\cal N}=1$ $SO$ and $Sp$ theories}

Using the brane configurations presented in the previous sections,
one can re-derive the ${\cal N}=1$ dualities for $SO$ and $Sp$ groups of
Refs.~\cite{Seiberg,IntSeib,IntPoul}. In the context of string theory these
dualities have been obtained first in Ref.~\cite{Nick} with the
use of an orientifold four plane.

\begin{figure}
\PSbox{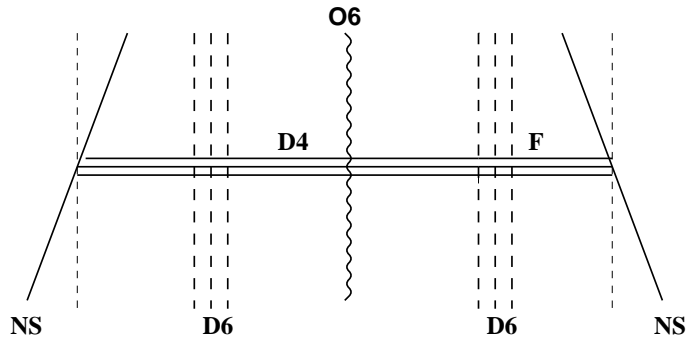 hscale=70 vscale=70 hoffset=100
voffset=0}{13.7cm}{4.5cm}
\caption{The brane set-up of the electric $SO$/$Sp$ theory with $F$ flavors.
\label{duality1}}
\end{figure}

In order to obtain an ${\cal N}=1$ $SO$ or $Sp$ theory without tensors, we
consider the general case of Fig.~\ref{setup} with $\alpha$ not equal
to $0$ or $\pi/2$. In this case we have an ${\cal N}=1$ theory with a massive 
adjoint and a massive tensor. In order to obtain the flavors of $SO$ and 
$Sp$ we have to introduce $F$ additional D6 branes on both sides of the 
O6-plane parallel to it (Fig.~\ref{duality1}). 
The resulting theory is ${\cal N}=1$ 
$SO$ or $Sp$ with $F$ flavors
($2F$ vectors of $SO$ or $2F$ fundamentals of $Sp$). However, the 
presence of the massive tensors induces a tree-level superpotential 
term for the flavors. This can be seen by considering the ${\cal N}=2$ 
configuration 
of Fig.~\ref{sixbranes}. This theory has a tree-level superpotential of the
form 
\beq 
W_{{\cal N}=2}= QA\bar{Q},
\eeq
where $A$ is the adjoint of $SO$/$Sp$ while the $Q$, $\bar{Q}$ are the 
flavors forming ${\cal N}=2$ hypermultiplets. Rotating the two NS5 branes
generates a mass for the adjoint (the massive tensor appears
as well, but since it does not couple to the flavors its
presence does not leave an imprint on the moduli space, and we will
ignore it in 
this section). Thus the full superpotential of the ${\cal N}=1$ theory is
\beq
W=QA\bar{Q}+\mu A^2;
\eeq
integrating out the massive adjoint we get a superpotential 
\beq 
W\sim (Q\bar{Q})^2.
\eeq
One can easily check that the counting of the dimension of the moduli space 
from the brane picture agrees with the field theory 
result. The simplest way to see this is to realize that the counting 
of the moduli space in the brane picture as well as in the field theory
is exactly the same as for the ${\cal N}=2$ theories discussed
in the previous section.

\begin{figure}
\PSbox{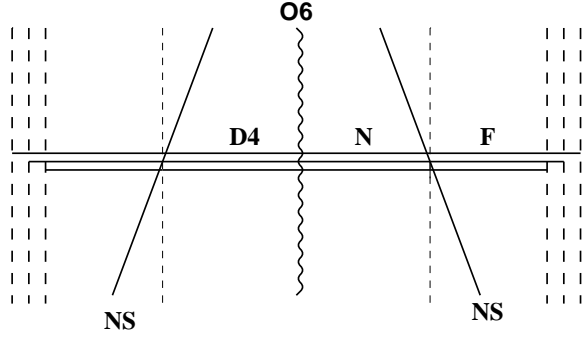 hscale=70 vscale=70 hoffset=100
voffset=0}{13.7cm}{4.5cm}
\caption{The brane set-up obtained after moving the D6 branes through the
NS5 branes. \label{duality2}}
\end{figure}

In order to obtain the dual description, we will move the D6 and the NS5
branes through each other, and use the linking number argument of
Ref.~\cite{HananyWitten} to obtain the number of D4 branes
created. The linking numbers for the NS5 branes are given by
\beq
L_5=\frac{1}{2} (n_{6L}-n_{6R})+n_{4R}-n_{4L},
\eeq
where $n_{6L,R}$ are the
D6 branes to the left or right of the NS5 branes,
and similarly $n_{4R,L}$ are the D4 branes to the right or left of the
NS5 branes. The linking numbers of the D6 branes are
\beq
L_6=\frac{1}{2} (n_{5L}-n_{5R})+n_{4R}-n_{4L},
\eeq
where $n_{5L,R}$ are the NS5 branes to the left or right of the
D6 brane, while  $n_{4R,L}$ are the D4 branes to the right or left of
the D6 brane. We assume that for the linking number argument the O6-plane
can be treated as $\pm 4$ D6 branes.

First, we move the D6 branes on the right all the way to the right
(and their mirrors all the way to the left) as shown in Fig.~\ref{duality2}.
Considering the linking number $L_6$ of a given D6 brane, we can see that
that there has to be one additional D4 brane ending on it after the move
to conserve $L_6$. This is because originally $L_6=-\frac{1}{2}$,
while after the move it is $L_6'=\frac{1}{2}-n_{4L}$, thus $n_{4L}$ has to be
one. There is one additional D4 brane created for every D6 brane, altogether
there are $F$ D4 branes created on each side of the O6.

\begin{figure}
\PSbox{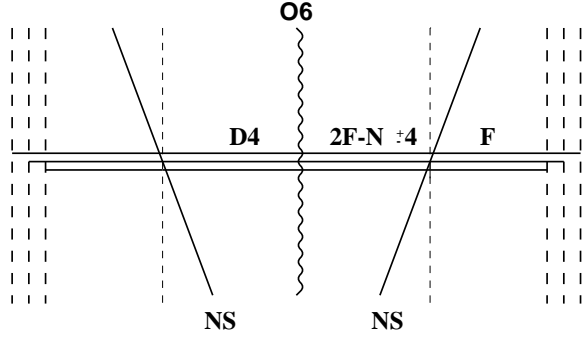 hscale=70 vscale=70 hoffset=100
voffset=0}{13.7cm}{4.5cm}
\caption{The brane set-up obtained after moving the NS5 branes through 
each other and the O6. \label{duality3}}
\end{figure}

\begin{figure}
\PSbox{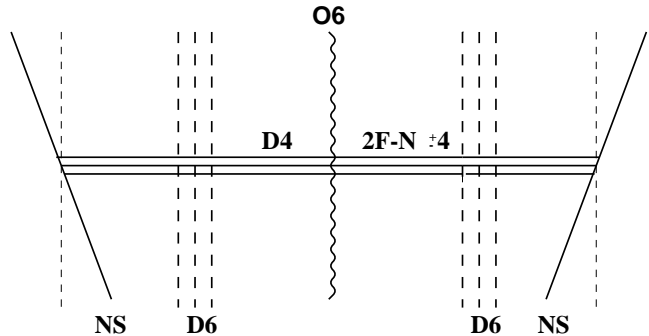 hscale=70 vscale=70 hoffset=100
voffset=0}{13.7cm}{4.5cm}
\caption{The brane set-up obtained after moving the D6 branes back through the
NS5 branes. \label{duality4}}
\end{figure}

Next, consider moving the NS5 branes through each other and through the
O6-plane. The resulting set-up for the magnetic theory is depicted in
Fig.~\ref{duality3}. To determine the number of D4 branes stretched
between the two NS5 branes we consider conservation of $L_5$
linking number. The linking number of the NS brane on the left is
originally $L_5=-\frac{1}{2}(\pm 4) +N-F$, since there are $N$ D4 branes
on the right, $F$ D4 branes on the left and $\pm 4$ D6 branes which
correspond to the orientifold on the
right. In the final state of Fig.~\ref{duality3} there are $\pm 4$
D6 branes on the left of the NS brane, $F$ D4 branes on the right and
$\tilde{N}$ D4 branes on the left, thus $L_5'= \frac{1}{2}(\pm 4) +F-
\tilde{N}$, from which we obtain $\tilde{N}=2F-N\pm 4$ which is the correct
size of the dual gauge group. Furthermore, we can move the D6 branes 
through the NS5 branes back again, resulting in the configuration of
Fig.~\ref{duality4}. This is just like the configuration of the electric 
theory with a different gauge group, which does correspond to
$SO(2F-N+4)$ or $Sp(2F-N-4)$ with $F$ flavors and a superpotential
\beq
W\sim (q\bar{q})^2,
\eeq
where $q$,$\bar{q}$ are the dual quarks of $SO(2F-N+4)$/$Sp(2F-N-4)$.
This is in agreement with the field theory result, since from the
${\cal N}=1$ duality we expect to get the dual gauge group $SO(2F-N+4)$
or $Sp(2F-N-4)$ and a superpotential term
\beq
W_{dual}= M^2 +Mq\bar{q},
\eeq
where the first term comes from the $(Q\bar{Q})^2$ tree-level
superpotential of the electric theory, while the $Mq\bar{q}$ term is the 
additional superpotential required for duality. Thus the mesons are massive 
and one can integrate them out, leaving the superpotential
\beq
W_{dual}\sim (q\bar{q})^2.
\eeq
This is exactly what we find in the brane picture of Fig.~\ref{duality4}. 

\subsection{Duality of $U(N)$ with a Symmetric or Antisymmetric Flavor}

\begin{figure}
\PSbox{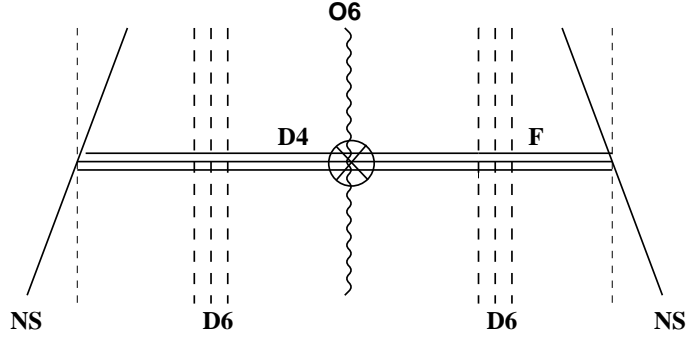 hscale=70 vscale=70 hoffset=100
voffset=0}{13.7cm}{4.5cm}
\caption{The brane configuration corresponding to the 
electric $SU(N)$ theory with a symmetric or an
antisymmetric flavor and $F$ fundamental flavors. The $\otimes$ denotes 
an NS5 brane perpendicular to the O6.\label{SUduality1}}
\end{figure}

Let us present one more example of duality, which takes advantage
of another brane configuration described previously. In Section~\ref{SUsection}
we outlined a configuration that gives a $U(N)$ theory with a flavor
of symmetric (antisymmetric) tensors. To that configuration we add
$F$ D6 branes as illustrated in Fig.~\ref{SUduality1}. These additional
six branes provide $F$ flavors in the fundamental representation.
As we already described in Section~\ref{SUsection} this theory has a tree-level
superpotential which includes the term $ (T \bar{T})^2$, 
where $T$ is a symmetric
or an antisymmetric tensor depending on the charge of the orientifold
plane. This is the theory considered in Sections 3 and 4 of
Ref.~\cite{ILS}, where duality was demonstrated after addition of the
superpotential term $W \sim (T \bar{T})^{k+1}$.
The field theory corresponding to our brane configuration has
additional superpotential terms which act as perturbations.
We discuss the (important) effect of these superpotential terms
in the field theory at the end of this section.  

\begin{figure}
\PSbox{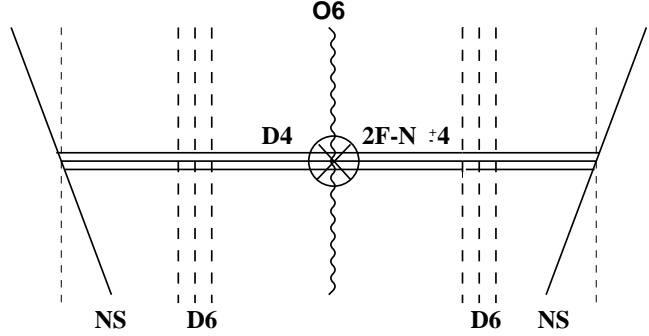 hscale=70 vscale=70 hoffset=100
voffset=0}{13.7cm}{4.5cm}
\caption{The magnetic configuration obtained after moving the 
NS5 branes through the D6 branes and each other \label{SUduality4}}
\end{figure}

Having already dealt with duality in a very similar set-up 
we can immediately derive the dual for this example.
We keep the D6 branes fixed and move only the outer NS5 branes.
First,  we move them closer to the orientifold, passing them through
the D6 branes. Then, we move the
NS5 branes across each other and the O6 and the NS5 brane in the center,
and finally move them outside the D6 branes. This final configuration
is illustrated in Fig.~\ref{SUduality4}. 
The number of D4 branes
created by this motion is identical to the one for $SO$($Sp$) duality with
fundamental flavors. For example, this can be seen by determining the
linking numbers of the various branes before and after the moves.
Thus after the move there are $2 F - N \pm 4$ D4 branes stretching between
the NS branes. The configuration is similar to our starting set-up, except for
a different number of D4 branes; it describes a theory
with $SU(2 F - N \pm 4)$ gauge group with a superpotential of the same form
as the one in the electric theory.

This result requires an explanation. The electric theory was constructed
by rotating the outer two NS5
branes of the ${\cal N}=2$ theory, that is by giving a mass to the 
adjoint superfield.
The resulting ${\cal N}=1$ theory with a massive adjoint has a 
superpotential 
\begin{displaymath}
  W=\mu A^2 + T A \bar T + Q A \bar{Q},
\end{displaymath}
and after integrating out the adjoint, the superpotential is
\begin{displaymath}
  W\sim (T \bar{T})^2 + (Q \bar{Q})^2 + Q \bar{T} T\bar{Q}.
\end{displaymath}
The dual which we read off of our brane construction is thus
an $SU(2 F -N\pm 4)$ gauge theory with $F$ flavors, a symmetric
(or antisymmetric) tensor flavor, and the superpotential
\begin{displaymath}
  W\sim (\tilde t \bar{\tilde t})^2 + (\tilde q \bar{\tilde q})^2 +
\tilde q \bar{\tilde t} \tilde t\bar{\tilde q}.
\end{displaymath}
We now proceed to compare this result with the expected field theory answer.
The field theory result can be obtained by using the dual presented
in Ref.~\cite{ILS} for the theory with the superpotential $W=(T \bar{T})^2$
and adding the two terms $(Q \bar{Q})^2 + Q \bar{T} T\bar{Q}$ as perturbations.
The chiral ring of the theory without perturbation 
consists of two meson operators $M_0=Q \bar{Q}$,
$M_1=Q \bar{T} T\bar{Q}$ and two tensors of the flavor group
$P=Q \bar{T} Q$, $\bar{P}= \bar{Q} T \bar{Q}$. The dual is an
$SU(3 F -N\pm 4)$ gauge theory with fundamental meson fields $P$,
$\bar{P}$, $M_0$, $M_1$, dual quarks $q,\bar{q}$,
dual tensors $t,\bar{t}$ and the superpotential
\begin{displaymath}
  W \sim (t \bar{t})^2 + M_1 \, q \bar{q} + M_0 \, q \bar{t} t \bar{q} +
     P \, q \bar{t} q + \bar{P} \, \bar{q} t \bar{q} + M_0^2 + M_1 .
\end{displaymath}
The last two terms correspond to the perturbations present in our 
brane set up.
We see that the meson field $M_0$ is massive and should be integrated out,
leading to a $(q \bar{t} t \bar{q})^2$ term in the superpotential.
Furthermore, the linear term $M_1$ forces
VEVs for the dual quarks $q$, $\bar{q}$. Since $M_1$ is an $F$ by $F$ matrix
in flavor space, all dual quarks get VEVs. They break the magnetic
gauge group $SU(3F-N \pm 4)$ to $SU(2 F-N \pm 4)$.
The dual quarks are eaten by the Higgs mechanism, while the tensors
decompose into tensors of $SU(2 F-N \pm 4)$, $F$ flavors and singlets.
The singlets get masses with the $P$ and $\bar{P}$ fields from the
$P q \bar{t} q + \bar{P} \bar{q} t \bar{q}$ terms. The remaining two terms
in the superpotential yield $W=(\tilde{t} \bar{\tilde{t}})^2 +
(\tilde{q} \bar{\tilde{q}})^2 + \tilde{q} \bar{\tilde{t}} \tilde{t}
\bar{\tilde{q}} $, where these are fields coming from the decomposition
of the tensor fields $t$ and $\bar{t}$. This is the same
superpotential and gauge group which we obtained from the brane move.


\section{M-theory}
\setcounter{equation}{0}

\subsection{$SO(2N)$}
We now proceed to construct new M-theory curves for the $SO(2N)$ theory.
(The extension of all these results to $SO(2N+1)$ is trivial.)
The construction includes an O6-plane which is at a point in $v$
and extends in the $w$ direction and $x_9$. There are two
NS5 branes, one on each side (in $x_6$) of the O6-plane 
(they are mirror images)
extending in the $w_\pm$ directions
where
\beq
w_\pm = w \pm \mu v ~.
\eeq

\begin{figure}
\PSbox{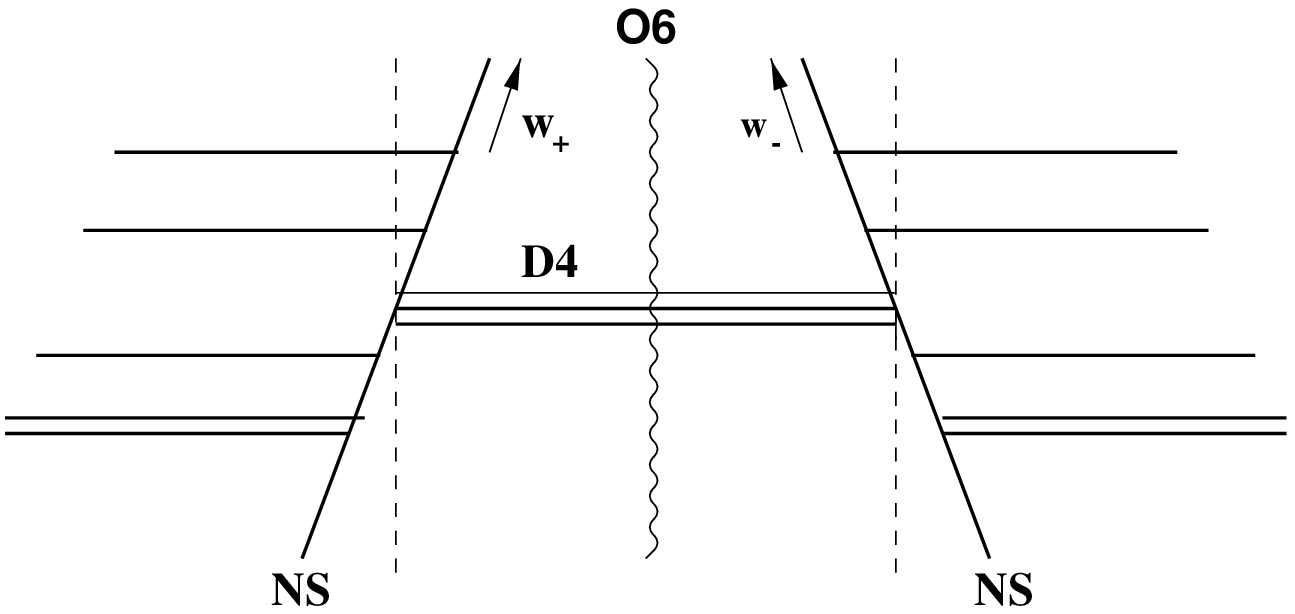 hscale=70 vscale=70 hoffset=100
voffset=0}{13.7cm}{4.5cm}
\caption{The brane set-up corresponding to ${\cal N}=1$ $SO$/$Sp$ theories
with massive flavors from semi-infinite D4 branes.\label{semiinfinite}}
\end{figure}

The NS5 branes have $2N$ D4 branes stretched between them, and $F$ 
semi-infinite
D4 branes extending to the left (negative $x_6$ direction) from the left
NS5 brane, and  mirrors of these on the right. This set-up is summarized
in Fig.~\ref{semiinfinite},  and it describes an $SO(2N)$ theory with 
$F$ flavors, that is $2 F$ vectors.
To describe the orientifold we note that it carries D6 brane charge
four, and assume that its effects can be described in the same fashion
as four coincident D6 branes. A D6 brane of type IIA string theory
is described by Taub-NUT space in M theory. For our purposes it is
sufficient to use one of the complex structures of this space and
then the orientifold is given by \cite{LL}:
\beq
x y = v^4 ~,
\label{SOxy}
\eeq
where $y \sim t = \exp(-s/R)$ for large $y$ and $x \sim t^{-1}$ for large
$x$. Here $s=x_6 + i x_{10}$.
To describe the curve in this space we need two equations in the
complex variables $v$,
$w$, $x$ and $y$.  We will simply present the curves, and then give evidence
that they are correct:
\beq
w_+ w_- = \mu \xi ~,
\label{SOwpm}\eeq
and
\beq
\left(\mu \xi \right)^{N-1} y \prod_{i=1}^F\left( {{w_+ - m_i}\over{w_- - m_i}}
\right) = v^2 w_+^{2N-2} ~,
\label{SOyw}
\eeq
where
\beq
\xi =  \left( \Lambda_{N,F}^{3(N-1) - F} \prod_i^F m_i  \right)^{{1}\over{N-1}}
{}~.
\eeq

It is straightforward to check that equations (\ref{SOxy}), (\ref{SOwpm}), and
(\ref{SOyw}) are invariant under the orientifold projection which exchanges:
\beq
 x & \leftrightarrow & y \nonumber \\
v & \leftrightarrow & -v \\
w  & \leftrightarrow & w \nonumber
\eeq
which imply $w_+ \leftrightarrow w_-$. Furthermore the curve is invariant
under $R_v$ and $R_w$ as defined in Table 1.

\begin{table}[htbp]
\centering
\begin{tabular}{ccc}
  & $R_v$ & $R_w$ \\
 \Yasymm & 2 &0 $\vbr$ \\
 \Ysymm & 0 &2 $\vbr$ \\
$\mu$ & -2 & 2 $\vbr$ \\
$m_i$ & 0 & 2 $\vbr$ \\
$Q$ & 1 & 0 $\vbr$ \\
$M$ & 2 & 0 $\vbr$ \\
$\Lambda_{N,F}^{3(N-1) - F}$ & $2(N-1)$ & $2(N-1) -2 F$ $\vbr$ \\
$\xi$ & 2 &  2 $\vbr$ \\
\end{tabular}
\label{Sp}
\parbox{4in}{\caption{R-charges for $SO(2N)$.}}
\end{table}

The most convincing evidence that the curves are correct is seen by
taking the string theory limit, as we describe in detail in the next
subsection.
The string theory limit is achieved by taking $R \rightarrow 0$, where $R$
is the radius of the compact 11-th dimension.  Taking this limit we see that
the M-theory curve reduces to the string theory picture of 5 branes and
4 branes described above. Readers
who are already convinced that the curves are correct may skip ahead.

\subsection{The String Theory Limit}
We now consider the string theory limit of the curve.
We take $R \rightarrow 0$ with
\beq
\Lambda_{N,F}^{3(N-1) - F}  \sim e^{-L/R}  ~,
\label{L}
\eeq
while holding  $L$ and $m_i$ fixed.
In order to see what the M theory configuration reduces to we first
consider a point on the curve with fixed large $w_+ \gg m_i$, and search for 
solutions
with large $y \sim t = e^{-s/R}$ (here $s < 0$).
Equation (\ref{SOwpm}) implies that
\beq
w_- \rightarrow 0 ~,
\eeq
while equation
(\ref{SOyw}) implies that
\beq
e^{-L/R} e^{-s/R} \sim {\rm const.} \,\,w_+ ^{2(N-1)-F} ~,
\eeq
so $s = -L$ in the limit $R \rightarrow 0$.
These solutions correspond to the NS5 brane on the left. Next consider
\beq
w_+ \sim m_i + e^{-c/R}~.
\eeq
Equation
(\ref{SOyw}) implies that
\beq
e^{-L/R} e^{-s/R} e^{-c/R} \sim {\rm const.}  ~,
\eeq
so $s = -L-c$. Thus we see $F$ solutions (one for each $m_i$) extending to the
left (negative $x_6$) that approach $w_+ = m_i$ as $s \rightarrow - \infty$.
These are precisely the $F$ semi-infinite D4 branes on the left.  The mirror
solutions can be found on the right for large $x$ (instead of large $y$).

Finally consider \beq
w_+ \sim  e^{-c/R}~.
\eeq
Equation
(\ref{SOyw}) implies that for $2c < L/(N-1)$
\beq
e^{-L/R} e^{-s/R} \sim {\rm const.} \,\, e^{- 2N c/R}   ~,
\eeq
so $s = -L+ 2Nc$ and these solutions sit to the right of the NS5 brane
and are the color D4 branes.

In M-theory the  NS5 branes can bend in
the $x_6$ direction, corresponding to the running coupling of the field
theory. We can also check
that our curve correctly reproduces this behavior. Taking a slice of the
brane-surface
with large, fixed $y$ we should find that the
number of solutions of the curve corresponds precisely to the number of branes
in string
picture.  To see this we rewrite equation
(\ref{SOyw}) in terms of $y$ and $w_+$. Using equation (\ref{SOwpm}) and
\beq
v = {{1}\over{2 \mu}} \left( w_+ - {{\mu \xi}\over{w_+}} \right)
\label{v}
\eeq
we find:
\beq
\left(\mu \xi \right)^{N-1} y \prod_{i=1}^F \left(w_+ - m_i \right)
w_+^{F-2N+4} = \left(w_+^2 - \mu \xi \right)^2  \prod_{i=1}^F \left( \mu \xi -
m_i w_+ \right)
\label{SOtubes}
\eeq
There are three cases:
a) $F \ge 2N$, there are $2F-2N+4$ solutions, $F$ with $w_+ \approx m_i$ (the
``flavor-branes") and
$F-2N+4$ with $w_+ \approx 0$ (this corresponds to the NS5 brane on the right
bending to the left);
b) $2N-4 <F < 2N$, there are $F+4$ solutions, $F$ with $w_+ \approx m_i$,
$2N-F$
with large $w_+$ (the NS5 brane on the left bends to the left), and $F-2N+4$
solutions with
$w_+ \approx 0$ (these solutions are related to the O6-plane, and in the string
theory
sit to the right of the NS5 brane on the left as describe above);
c) $F \le 2N-4$, there are $2N$ solutions, $F$ with $w_+ \approx m_i$, and
$2N-F$
with large $w_+$ (again the NS5 brane on the left bends to the left).

\subsection{Decoupling a Flavor}
It is also a simple exercise to check that decoupling a flavor works correctly.
For example take $m_F \rightarrow \infty$, this has the effect of replacing
\beq
\prod_{i=1}^F \rightarrow \prod_{i=1}^{F-1}
\eeq
and making the identification (from one-loop matching of the gauge coupling)
\beq
\Lambda_{N,F}^{3(N-1) - F} \prod_i^F m_i = \Lambda_{N,F-1}^{3(N-1) - (F-1)}
\prod_i^{F-1} m_i ~.
\eeq

\subsection{Duality}
We also see that the M-theory curve correctly describes the dual gauge
theory by taking the string theory limit ($R \rightarrow 0$)
with $\Lambda_{N,F} \rightarrow
\infty$ (for the asymptotically free case)
rather than $\Lambda_{N,F}
\rightarrow 0$ as we did above. Equation (\ref{L}) then implies that $L <0$.
The intrinsic scale, $\tilde \Lambda_{\tilde N,F}$, of the dual theory is
related to $\Lambda_{N,F}$ by:
\beq
\Lambda_{N,F}^b \tilde \Lambda_{\tilde N,F}^{\tilde b} = \mu^{b + \tilde b}~,
\eeq
where $b$ and $\tilde b$ are the one-loop $\beta$-function coefficients,
and $\mu$ is a scale which is approximately equal to the string
scale by dimensional analysis.
Thus, we see that the limit $\Lambda_{N,F} \rightarrow \infty$ corresponds
to\footnote{We are stating this for the case
of asymptotically free electric and magnetic theories ($b,\tilde{b} > 0$),
the extension to the other cases
($b \le 0$ or $\tilde{b}\le 0$) is straightforward and gives further
evidence for duality. For details see Ref.~\cite{MR}.}
 $\tilde \Lambda_{\tilde N,F} \rightarrow 0$. To make things simpler we
rewrite equation (\ref{SOyw})  with $w_-$ expressed in terms of $w_+$ and
vice versa by use of Equation (\ref{SOwpm}). More explicitly:
\beq
\left(\mu \xi \right)^{N-1} y \prod_{i=1}^F\left( {{ {{\mu \xi}\over{w_-}} -
m_i}\over{ {{\mu \xi}\over{w_+}} - m_i}} \right) = v^2 \left( {{\mu
\xi}\over{w_-}} \right)^{2N-2}
\label{SOywsub}
\eeq
is equivalent to
\beq
\left(\mu \xi \right)^{\tilde N-1} y \prod_{i=1}^F\left( {{w_- - \mu
M_i}\over{w_+ - \mu M_i}} \right) = v^2 w_-^{2 \tilde N-2} ~,
\label{SOywdual}
\eeq
where
\beq
\tilde N = F- N +2 ~,
\eeq
corresponding to the dual gauge group $SO(\tilde N)$,
and $M_i$ is the meson VEV:
\beq
M_i = {{\xi}\over{m_i}} .
\eeq
Note that
\beq
\xi =
\left( \tilde \Lambda_{\tilde N,F}^{3(\tilde N-1) - F}  \prod_i^F m_i
\right)^{{1}\over{\tilde N-1}}
\eeq
In order to hold the meson VEV fixed in the $R \rightarrow 0$ limit
we must take:
\beq
m_i^{F-N+1} \sim e^{L/R} \rightarrow 0 ~.
\eeq
Thus we see that the string theory limit of the M-theory curve with
$\Lambda_{N,F} \rightarrow \infty$ (i.e. $L < 0$) gives precisely the
string theory picture of the dual gauge theory.

\subsection{$Sp(2N)$}
Next we consider the M-theory curves for the $Sp(2N)$ theory.
The construction again includes an O6-plane (but with the opposite charge)
at a point in $v$ and
extending in the $w$ direction and $x_9$. There are two 
NS5 branes, one on each side (in $x_6$) of the O6-plane
extending in the $w_\pm$ directions.
The NS5 branes have $2N$ D4 branes stretched between them, and $F$ 
semi-infinite
D4 branes extending to the left from the left
NS5 brane, and  mirrors of these on the right (Fig.~\ref{semiinfinite}).
Again we need three equations in the complex variables $v$,
$w$, $x$ and $y$. As before we give the curves and then the evidence that they
are
correct.
The O6-plane is described by \cite{LL}:
\beq
x y = v^{-4} ~.
\label{Spxy}
\eeq
We also have
\beq
w_+ w_- = \mu \xi ~,
\label{Spwpm}\eeq
and
\beq
\left(\mu \xi \right)^{N+1} y \prod_{i=1}^F\left( {{w_+ - m_i}\over{w_- - m_i}}
\right) = v^{-2} w_+^{2N+2} ~.
\label{Spyw}
\eeq
It is straightforward to check that equations (\ref{Spxy}), (\ref{Spwpm}), and
(\ref{Spyw}) satisfy the correct symmetries and to check that decoupling a
flavor works correctly.

\subsection{The String Limit}
We now consider the string theory limit $R \rightarrow 0$ of the curve with
\beq
\xi^{N+1 }  \sim e^{-L/R} ~,
\label{SpL}
\eeq
$L > 0$  and $m_i$ fixed.
We first consider  fixed $w_+ \gg m_i$,
and search for solutions
with large $y \sim t = e^{-s/R}$.  Equation (\ref{Spwpm}) implies that
\beq
w_- \rightarrow 0 ~,
\eeq
while equation
(\ref{Spyw}) implies that
\beq
e^{-L/R} e^{-s/R} \sim {\rm const.} \,\,w_+ ^{2N-F} ~,
\eeq
so $s = -L$ in the limit $R \rightarrow 0$.
These solutions correspond to the NS5 brane on the left. Next consider
\beq
w_+ \sim m_i + e^{-c/R}~.
\eeq
Equation
(\ref{SOyw}) implies that
\beq
e^{-L/R} e^{-s/R} e^{- c/R} \sim {\rm const.}  ~,
\eeq
so $s = -L- c$. Thus we see $F$ solutions (one for each $m_i$) extending to the
left that approach $w_+ = m_i$ as $s \rightarrow - \infty$.
These are the $F$ semi-infinite D4 branes on the left.  The mirror
solutions can be found on the right for large $x$.
Finally consider
\beq
w_+ \sim  e^{-c/R}~,
\eeq
equation
(\ref{Spyw}) implies that for $2c < L/(N+1)$
\beq
e^{-L/R} e^{-s/R} \sim {\rm const.} \,\, e^{- (2N +2) c/R}   ~,
\eeq
so $s = -L+ (2N +2) c$ and these solutions sit to the right of the NS5 brane;
they are the color D4 branes.

We can also check  the curve by counting the number of solutions for  large,
fixed
$y$ and comparing to the number of branes in the string picture.
Using equations (\ref{v}) and (\ref{Spwpm})
we rewrite equation (\ref{Spyw}) in terms of $y$ and $w_+$:
\beq
\left(\mu \xi \right)^{N+1} y \prod_{i=1}^F \left(w_+ - m_i \right)
w_+^{F-2N-4} \left(w_+^2 - \mu \xi \right)^2 =   \prod_{i=1}^F \left( \mu \xi -
m_i w_+ \right)
\label{Sptubes}
\eeq
There are again three cases:
a) $F \ge 2N+4$, there are $2F-2N$ solutions, $F$ with $w_+ \approx m_i$ (the
``flavor-branes"),  $F-2N-4$ with $w_+ \approx 0$ (this corresponds to the
NS5 brane on the right
bending to the left) and four solutions with $w_+^2 \approx \mu \xi$
(associated with the
O6-plane);
b) $2N+2  \le F < 2N+4$, there are $F+4$ solutions, $F$ with $w_+ \approx m_i$,
and four solutions with $w_+^2 \approx \mu \xi$ (associated with the O6-plane);
c) $F < 2N+2$, there are $2N+6$ solutions, $F$ with $w_+ \approx m_i$, and
$2N+2-F$
with large $w_+$ (the NS5 brane on the left bends to the left), and four
solutions with $w_+^2 \approx \mu \xi$ (associated with the O6-plane).

\subsection{Duality}

We can also see that the M-theory curve correctly describes the dual gauge
theory
by simply
rewriting Equation (\ref{Spyw}) and taking the limit $R \rightarrow 0$ with
$\Lambda \rightarrow \infty$.
The curve,
\beq
\left(\mu \xi \right)^{N+1} y \prod_{i=1}^F\left( {{ {{\mu \xi}\over{w_-}} -
m_i}\over{ {{\mu \xi}\over{w_+}} - m_i}} \right) = v^{-2} \left( {{\mu
\xi}\over{w_-}} \right)^{2N+2} ~,
\label{Spywsub}
\eeq
is equivalent to
\beq
\left(\mu \xi \right)^{\tilde N+1} y \prod_{i=1}^F\left( {{w_- - \mu
M_i}\over{w_+ - \mu M_i}} \right) = v^{-2} w_-^{2 \tilde N+2} ~,
\label{Spywdual}
\eeq
where
\beq
\tilde N = F- N - 2 ~,
\eeq
corresponding to the dual gauge group $Sp(2 \tilde N)$.
Again we see that this correctly reproduces dual gauge theory in the
string theory limit.


\section{Conclusions}
We have investigated brane realizations of ${\cal N}=1$ $SO$ and $Sp$ gauge  
theories with two-index tensor representations. Our construction
employed an orientifold six plane. By studying various limiting
cases we have established the field content and the tree-level
superpotential of our brane construction. The important limiting cases
correspond to ${\cal N}=2$ theory with tensor hypermultiplets and
${\cal N}=1$ theories with massless tensors and no superpotential.
For these cases, we have checked that the brane degrees of
freedom are in exact correspondence
with the flat directions in field theory.

Consistency requirements impose two new restrictions
on brane configurations involving an orientifold six plane. First,
orientifolds with the $Sp$-type projection can be only crossed by pairs
of D4 branes. A single D4 brane is not invariant under
the projection of $Sp$-type. Second, the s-rule needs to be generalized
to our s'-rule in the presence of an orientifold six plane
with negative charge. Our first
observation says that D4 branes crossing an O6-plane
have to group into pairs invariant
under the orientifold projection. Then, according to the s'-rule
only one of the two branes in each
such pair can end on a given D6 brane parallel to an O6-plane.

We have also studied duality for $SO$ and $Sp$ groups both in the string
theory framework and in M-theory. In string theory, we included D6 branes
which transform under flavor symmetries. The dual gauge group and meson
excitation are identified when the original brane configuration is
deformed by moving NS5 and D6 branes. In M-theory, we included semi-infinite
branes and found the equations describing the single brane configuration.
This configuration encodes the information about the dual
gauge group and the vacuum expectation values of the meson fields.

\section*{Acknowledgments}
We are grateful to Jan de Boer, Kentaro Hori, Hirosi Ooguri,
Yaron Oz, Erich Poppitz and Raman Sundrum for useful discussions.
We also thank Jan de Boer, Yaron Oz and Zheng Yin for comments on the 
manuscript.
C.C. is a research fellow of the Miller Institute for Basic Research
in Science. C.C. and J.T. are supported in part by the National Science
Foundation under grant PHY-95-14797, and are also partially supported by
the Department of Energy under contract DE-AC03-76SF00098.
M.S. is supported  by the U.S.
Department of Energy under grant \#DE-FG02-91ER40676.
W.S. is supported by the Department of Energy
under contract DE-FG03-97ER405046.

\end{document}